\documentstyle[12pt,epsfig]{article}
\newcommand{\eqref}[1]{(\ref{#1})}              % equation numbers
\newcommand{\eqcm}{\; ,}
\newcommand{\GeV}{~{\rm GeV}}                   % units in math mode
\newcommand{\obs}{{\cal O}}                     % Observable
\newcommand{\weinberg}{\sin^2\! \theta_W}       % Weinberg angle

\begin{document}
\begin{flushright}
  DAMTP-96-25 \\
\end{flushright}

\vspace{0.5cm}

\begin{center}
  {\Large \bf Optimal observables for measuring three-gauge-boson
    couplings in $e^+ e^- \to W^+ W^-$}

\vspace{1.5cm}

M.~Diehl \\
{\em DAMTP, Silver Street, Cambridge CB3 9EW, England\/}
\\[\baselineskip]

and \\[\baselineskip]

O.~Nachtmann \\
{\em Institut f\"ur Theoretische Physik, Philosophenweg 16, 69120
  Heidelberg, Germany}

\vspace{2.5cm}

\parbox{0.9\textwidth}{Observables with optimised sensitivity to the $WWZ$ and
  $WW\gamma$-couplings in $e^+ e^- \to W^+ W^-$ are investigated. They
  allow for separate studies of CP violation and of absorptive parts
  in the amplitude. We have calculated expected statistical errors on
  extracted couplings at $\sqrt{s} = 500 \GeV$, with unpolarised or
  longitudinally polarised beams.}

\vspace{\fill}

{\em Talk given at the Workshop on Physics with $e^+ e^-$ Linear
  Colliders, \\ Gran Sasso, Italy, 2--3 June 1995 \/}

\end{center}

\newpage

\baselineskip=18pt

% TEXT BEGINS HERE

\section{Introduction}

In this contribution we propose a method to measure the trilinear
gauge couplings in the reaction $e^+ e^- \to W^+ W^-$. For the $WWZ$
and $WW\gamma$ vertices we use the most general parametrisation
consistent with Lorentz invariance of Hagiwara et
al.~\cite{md:Hagiwara}. It describes each vertex by seven form factors
$f_i$, three of which correspond to CP violating couplings.

We shall consider the decay channels $W\, W \to \ell\, \nu_\ell + {\rm
  jet}\,\, {\rm jet}$, where $\ell$ is a light lepton $e$ or $\mu$ and
the jets originate from a quark-antiquark pair.  For these channels a
complete kinematical reconstruction of the four-fermion final state is
possible, apart from the ambiguity to associate the jets with the
quark and the antiquark in case the jet charge cannot be measured.

In the following section we present a method to measure the form
factors $f_i$ with maximal statistical sensitivity, and in
sec.~\ref{md:results} we describe some numerical results for the
accuracy to be attained at a 500 GeV collider. Details can be found in
\cite{md:DiehlNachtmann}.

\section{The method of optimal observables}

Let us denote by $\phi$ the full set of reconstructed kinematical
variables, e.g.~the scattering angle of the $W^-$ in the c.m.~frame
and the decay angles of the $W$ in their respective rest frames.
Furthermore let $g_i$ be the difference between $f_i$ and its value in
the SM at tree level. Since the trilinear gauge couplings enter
linearly in the amplitude of our process the differential cross
section can be written as
\begin{equation}
  \frac{d\sigma}{d\phi} = S_0(\phi) + \sum_i S_{1,i}(\phi) \, g_i +
  \sum_{i,j} S_{2,ij}(\phi) \, g_i g_j \eqcm
  \label{md:crosssection}
\end{equation}
where it is understood that one has symmetrised over any kinematical
ambiguities. The idea of optimal observables is to measure the
distribution of the functions\footnote{The functions $\obs_i(\phi)$
  defined here are available as a FORTRAN routine from the authors.}
\begin{equation}
  \obs_i(\phi) = \frac{S_{1,i}(\phi)}{S_0(\phi)}
  \label{md:observables}
\end{equation}
and to determine the couplings from their mean values $\langle \obs_i
\rangle$. To first order\footnote{A leading order expansion in $g_i$
  is used here, assuming that deviations from the SM at tree level are
  small, but this is not essential for the method to work.} in the
$g_i$ one has
\begin{equation}
  \langle \obs_i \rangle = {\langle \obs_i \rangle}_0 + \sum_j
  c_{ij} \, g_j \eqcm
  \label{md:meanvalue}
\end{equation}
from which the $g_i$ can be extracted since ${\langle \obs_i
  \rangle}_0$ and $c_{ij}$ are calculable given \eqref{md:observables}
and \eqref{md:crosssection}. From the distribution of the $\obs_i$ one
also obtains the statistical errors on their mean values, and the
observables $\obs_i$ have been constructed to minimise the induced
statistical errors on the extracted couplings $g_i$ \cite{md:Atwood}.
More precisely, it can be shown \cite{md:DiehlNachtmann} that in the
limit of small $g_i$ they cannot be smaller in {\em any\/} other
method, including a fit to the full distribution of the variables
$\phi$.

To have tractable expressions for the observables, some approximations
such as taking the cross section~\eqref{md:crosssection} at tree level
will be necessary in practice. They need however not be made in the
extraction of the couplings: various theoretical and experimental
effects such as radiative corrections or detector resolution might be
taken into account when calculating the coefficients ${\langle \obs_i
  \rangle}_0$ and $c_{ij}$ in eq.~\eqref{md:meanvalue}.

This method is particularly well suited for testing discrete
symmetries, because for an observable which corresponds to a CP
violating coupling $g_i$, a nonzero mean ${\langle \obs_i \rangle}$ is
an unambiguous sign for CP violation in the reaction, provided
detector and data selection are CP blind.\footnote{Optimal observables
  have already been used to search for CP violation in $\tau$-pair
  production at LEP1, with a clear gain of sensitivity over non
  optimised observables \cite{md:LEP1}.} A similar statement holds for
the study of absorptive parts in the scattering amplitude.

\section{Numerical estimates for a 500 GeV collider} \label{md:results}

To estimate the statistical precision of our method we have calculated
the errors in measuring real and imaginary parts of the form factors
$f_i^Z$ and $f_i^\gamma$ of \cite{md:Hagiwara}. For an integrated
luminosity of 10 fb$^{-1}$ at $\sqrt{s} = 500\GeV$ and unpolarised
beams we find 1-$\sigma$ errors between $2 \cdot 10^{-4}$ and $2 \cdot
10^{-2}$.  With the additional information of the jet charge in each
event they decrease by a factor between 1.2 and 2.7.

For some of the form factors the errors are strongly correlated. This
holds in particular for pairs of $f_i^Z$ and $f_i^\gamma$, because due
to the interference of the $WWZ$ and $WW\gamma$ vertices they appear
as linear combinations in the amplitude. For left and right handed
incident electrons these are, respectively,
\begin{eqnarray}
  f_i^L &=& 4 \weinberg\, f_i^\gamma + (2 - 4 \weinberg)\,
  \frac{s}{s
    - M_Z^2}\, f_i^Z
  \nonumber \\
  f_i^R &=& 4 \weinberg\, f_i^\gamma - 4 \weinberg\, \frac{s}{s
    - M_Z^2}\, f_i^Z \eqcm
  \label{md:leftright}
\end{eqnarray}
where $\theta_W$ is the weak mixing angle. The correlations for pairs
$f_i^L$ and $f_i^R$ turn out to be much weaker. With the above
parameters the errors are between $2 \cdot 10^{-4}$ and $2 \cdot
10^{-2}$ for the $f_i^L$, and between $5 \cdot 10^{-4}$ and $7 \cdot
10^{-2}$ for the $f_i^R$. With beams of left handed $e^-$ or right
handed $e^+$ one would exclusively measure the $f_i^L$.  We find that
given the same number of events the statistical errors on the $f_i^L$
are almost the same with or without polarisation.

\end{document}